\documentclass[prd]{revtex4}
\usepackage{graphicx}
\usepackage{amssymb,amsmath}
\usepackage{enumerate}

\newcommand{\be}{\begin{eqnarray}}
\newcommand{\ee}{\end{eqnarray}}

\begin{document}

\title{Deterministic chaos in the large-scale universe: data versus speculations}

\author{A. Bershadskii}

\affiliation{
ICAR, P.O. Box 31155, Jerusalem 91000, Israel
}

\begin{abstract}

It is shown, that the both angular CMB Doppler spectrum: $C_l$ (Planck space telescope - cosmic microwave background \cite{Planck}) and the 3D galaxy-galaxy power spectrum: $P(k)$ (Sloan Digital Sky Survey SDSS-II \cite{teg1}), 
exhibit a considerable range with an exponential decay: $370 < l < 2500$ and 
$0.05 < k < 0.27~~(h/Mpc)$, respectively. The rates of the exponential decay are 
$l_c \simeq 300$ for $C_l \sim \exp-(l/l_c)$ and $k_c \simeq 0.09~~(h/Mpc)$ for $P(k) \sim \exp-(k/k_c)$. A waviness is observed along a straight line representing the exponential decay in the log-linear scale graphs of these spectra. In both cases the waviness has period (distance between peaks) equal to the same $l_c$ and $k_c$ as for the exponential decay. It means that the waviness is generated by the same, presumably chaotic, mechanism that generates the exponential decay. A more complex, distributed, chaos is observed in the Baryon Oscillation
Spectroscopic Survey (the largest component of the SDSS-III containing nearly one million galaxies) \cite{ander}. In this case the $P(k)$ spectrum is a weighted superposition of the exponentials. At assumption that dynamics of the dispersive waves, driving the distributed chaos, is dominated by the effects of a surface tension this weighted superposition of the exponentials is 
converged to a compact form of a stretched exponential $P(k) \sim \exp-(k/k_b)^{1/2}$, in good agreement with the data.

\end{abstract} 


\maketitle

\begin{flushright}
"Throwing pebbles into the water, look at the ripples they form on \\
the surface, otherwise, such occupation becomes an idle pastime."  \\

\~~~~~~~~~~~~~~~~~~~~~~~\\
Kozma Prutkov
\end{flushright}

\section{Introduction}

The phenomenon of large scale inhomogeneity in the Universe is one of the most fundamental phenomena in modern astrophysics. Despite (or may be because) of abundance of the high quality data our understanding of this phenomenon is now in a rather confused state. Early simple self-similar (scale invariant or scaling) picture was challenged by appearance of the data sets with different values of the scaling exponent. Moreover, advanced 3D analysis of 205,443 galaxies from the Sloan Digital Sky Survey (SDSS-II, mean redshift $ z \simeq 0.1$ ) data shows that: "The
power spectrum is not well-characterized by a single power law but unambiguously shows curvature." \cite{teg1}. The non-scaling character of the power spectra was also confirmed by the later observations in the Baryon Oscillation Spectroscopic Survey, the largest component of the third Sloan Digital Sky Survey (SDSS-III, $0.2 < z < 0.7$) containing nearly one million galaxies \cite{ander}. That means absence of the scale invariance in the large scales. The non-scaling implies existence of at least one chosen scale in the random like system of the galaxies with a broad band spectrum. Where this chosen scale come from? It is assumed at present time that the large scale inhomogeneity of the observable Universe has its seeds at the last scattering surface inhomogeneities, which manifest themselves in the tiny fluctuations observed in the cosmic microwave background radiation (CMBR).

The recent high resolution cosmic microwave background (CMB) radiation measurements \cite{Planck} provide a possibility for quantitative investigation of
the nontrivial processes taking place at the {\it decoupling} of the baryon-photon plasma (the last scatering surface) in the early universe (the Planck space telescope scanned the microwave and submillimetre sky continuously for more than four years between Aug. 2009 and Oct. 2013). 

Before the decoupling the CMB photons were tightly coupled to the baryons by photon 
scattering forming so-called baryon-photon fluid. At a certain stage of the universe 
expansion the energy of the CMB photons becomes insufficient to keep hydrogen ionized and a recombination
process becomes dominating. The recombination is a gradual process, 
but it serves as a trigger and a catalyst of the decoupling of the baryon-photon fluid. 
    The fluid (or plasma) is in a motion. What type of motion it is: laminar, chaotic or turbulent? Acoustic waves, motions that came from the previous history were active at the last scattering (see Ref. \cite{beck} and references therein for the cosmological chaotic scenarios). The decoupling process itself is also a source of the moving energy. A recent CMB estimation of the corresponding Reynolds number \cite{ber1} shows that it is possible that the motion is transitional from a laminar to turbulent. There are many ways for such transition. One of them is through chaos \cite{swinney1}.   For dynamical systems broad band exponential frequency spectra are strong indications of a chaotic motion \cite{f},\cite{sig}. For fluid motion this indication can be also valid \cite{swinney1}.  
    
Fig. 1 (adapted from Ref. \cite{mm}) shows (in the semi-log scales) a power spectrum of temperature fluctuations observed in a simple, two-mode simulation of relaxation of a temperature filament in magnetized plasma with unstable drift waves after onset of chaos. The first peak corresponds to the fundamental mode $f/f_c =1$ and the next peaks $ f/f_c=2,3,4$ correspond to harmonics of the coherent modes driving the deterministic chaos. The dashed line in Fig. 1 indicates the exponential decay of the spectrum $\exp-(f/f_c)$. The dotted lines indicate positions of the peaks. One can see that period of the waviness along the exponential decay (distance between peaks) is equal to the rate of the exponential decay. It is clear that the both quantities: the period and the rate, are generated by the same chaotic mechanism with the chosen scale equal to $f_c$. The data were obtained by a ’synthetic probe’ imitating the probes used in the real plasma experiments, where
"...time signals are typically measured at a fixed spatial location by probes and are a manifestation of the effects of spatially extended structures,
generated by deterministic chaos, sweeping past the probes." \cite{mm}. Therefore the frequency spectra are manifestation of the spatial spectra and the chosen frequency scale $f_c$ is a manifestation of a chosen spatial (wave number) scale $k_c$. It is significant for our consideration because we will analyse just spatial (wave number) spectra in the next sections. Namely, we will show that the Doppler CMB power spectrum $C_l$ (obtained by the Planck mission team \cite{Planck}) as well as the 3D power spectrum $P(k)$ of galaxies from the SDSS-II \cite{teg1} both exhibit the same properties. That indicates deterministic chaos as a possible common basis for the mechanisms generating the nonscaling spectra in the both cases. It also 
opens a way to understand more complex situations observed in the much more large Sky Surveys such as the Baryon Oscillation Spectroscopic Survey (SDSS-III) using a technique called baryon acoustic oscillation (BAO) to determine the distances to the galaxies. The distributed chaos observed in the last case is characterized by a stretched exponential power spectrum $P(k) \sim \exp-(k/k_b)^{1/2}$ that comes from a weighted superposition of the exponentials.

\section{Inhomogeneities in the cosmic microwave background}

So-called visibility function is used to describe 'optical' properties of the plasma at decoupling stage. 
A measure of the width of the visibility function around its (rather sharp) 
maximum can be used for quantitative characterization of the decoupling stage. This stage is also called as 
last scattering shell (surface). The last scattering probability turns out to be a narrow peak around a 
decoupling redshift. The visibility function is defined as 
$$
g(t) = n_e \sigma_T a(t) \exp  \left\{-\int_t^{t_0} n_e(t')\sigma_T a(t')dt'\right\}  \eqno{(1)}
$$
where $a(t)$ is the expansion factor and $\sigma_T$ is the Thomson cross section. 
The motion of the scatters imprints a temperature fluctuation,
$\delta T$, on the CMB through the Doppler effect 
$$
\frac{\delta T ({\bf n})}{T} \sim \int g(L){\bf n} \cdot {\bf v}
({\bf x})~dL  \eqno{(2})
$$
where ${\bf n}$ is the direction (the unit vector) on the sky and
${\bf v}$ is the velocity field of the electrons evaluated along
the line of sight, ${\bf x} = L{\bf n}$.

In the CMB literature it is used to expand the CMB temperature
fluctuations in spherical harmonics
$$
\frac{ \delta T }{T} = \sum_{lm} a_{lm} Y_{lm} (\Theta, \phi)  \eqno{(3)}
$$
For isotropic situation the angular power spectrum of the fluctuations, $C_{l}$, is defined as
$$ 
\langle a_{lm}a_{l'm'}^{*} \rangle = C_{l} \delta_{ll'}\delta{mm'} \eqno{(4)}. 
$$
The angular power spectrum is analogous to the power spectrum $P(k)$ of density perturbations.
\begin{figure}
\begin{center}
\includegraphics[width=12cm \vspace{-0.5cm}]{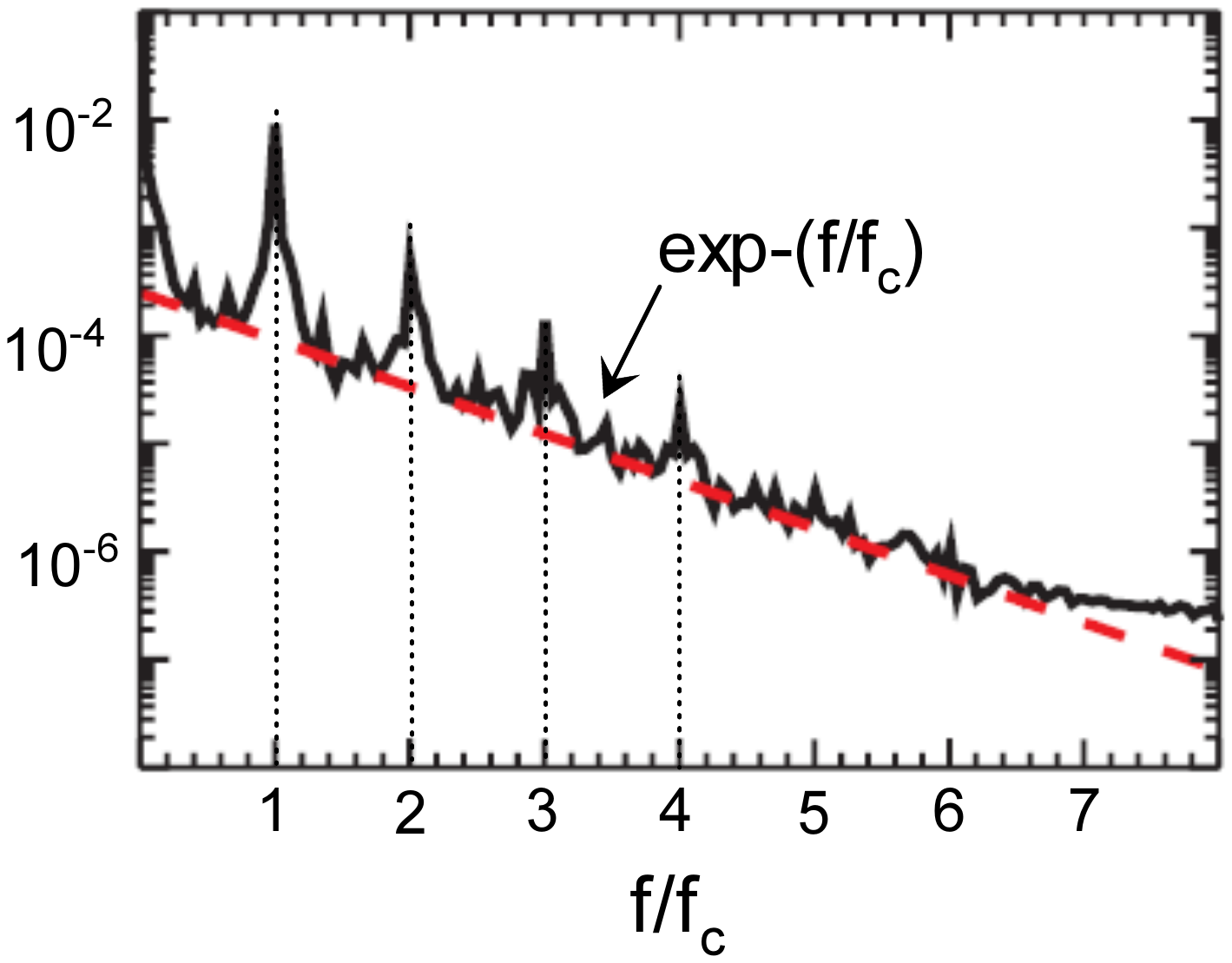}\vspace{-7cm}
\caption{\label{fig1}  The power spectrum of temperature fluctuations (in the semi-log scales) observed in a simple, two-mode simulation of relaxation of a temperature filament in magnetized plasmas after the onset of
chaos (adapted from the Ref. \cite{mm}). } 
\end{center}
\end{figure}

Figure 2 shows the angular spectrum $C_l$ calculated using the data obtained by 
the Planck space telescope. The data were taken from the file COM-PowerSpect-CMB-TT-hiL-binned-R2.01.txt on the site of the Planck Legacy Archive http://pla.esac.esa.int/pla.

The straight line is drawn in the Fig. 2 in order to indicate an exponential decay in the semi-log scales
$$
C_l \sim \exp -(l/l_c)   \eqno{(5)}
$$ 
with $l_c \simeq 300$. 

It should be noted that the waviness observed along the exponential decay has period (distance between peaks) equal to the same $l_c \simeq 300$. It means that the waviness is generated by the same, presumably chaotic, mechanism that produced the exponential decay (cf. Introduction and Fig. 1).

\begin{figure}
\begin{center}
\includegraphics[width=12cm \vspace{-1cm}]{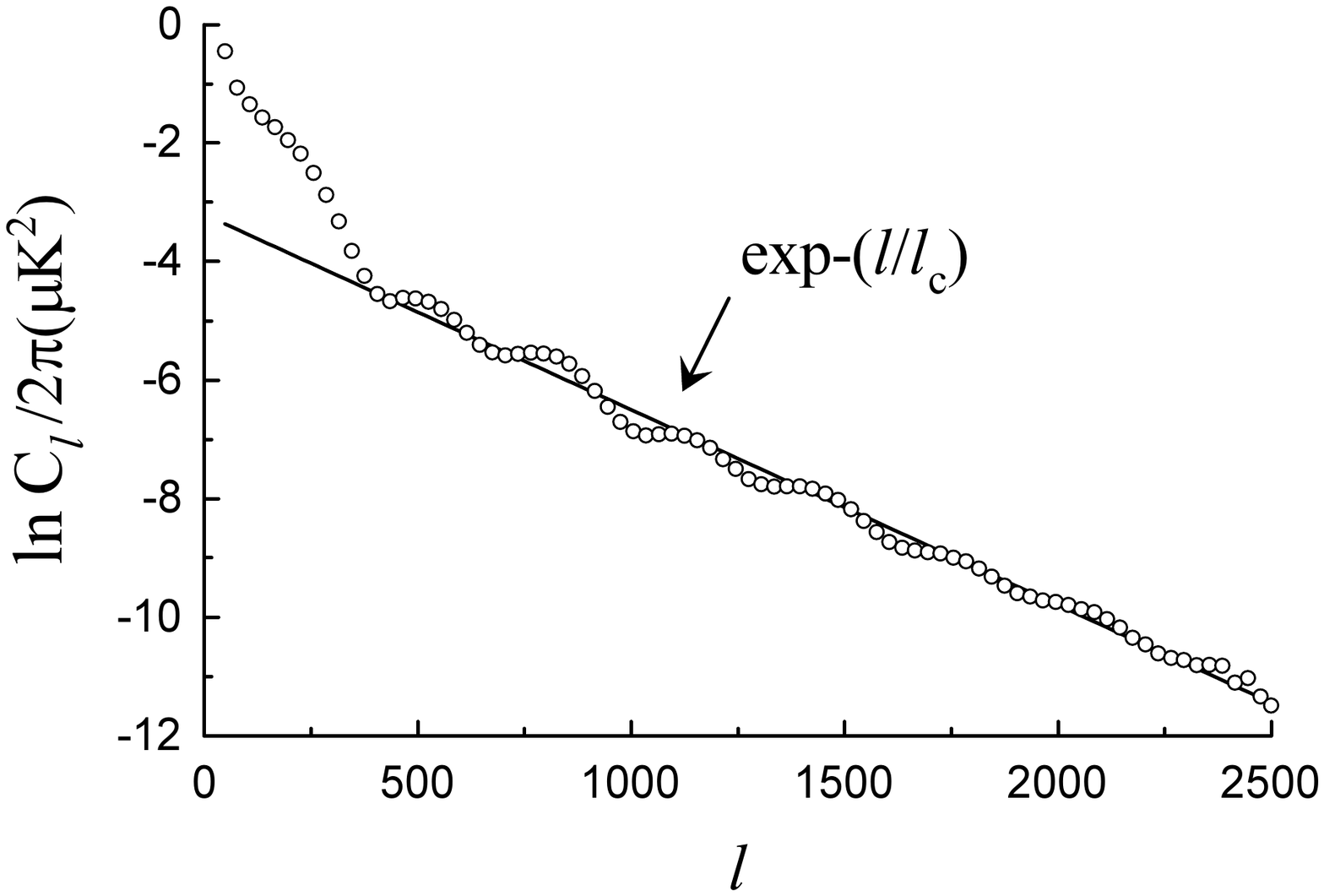}\vspace{-6cm}
\caption{\label{fig2} The binned angular spectrum $C_l$ (in the semi-log scales) calculated using the data obtained by the Planck space telescope. The straight line is drawn in order to indicate the exponential decay Eq. (5) with $l_c \simeq 300$.} 
\end{center}
\end{figure}

It was suggested in the CMB literature that on the scales $l > 2500$ (an 'improved' estimation $l > 1000)$ the spectrum is exponentially damped, due to the photon diffusion. The exponential spectrum Eq. (5), described in present note, takes place in the interval $370 < l < 2500$ and the period of the waviness is equal to the rate of the exponential decay $l_c$. Also the exponential damping mentioned in the literature is not actually exponential in the form of Eq. (5), because it is caused by a diffusion process and instead of $l/l_c$ in Eq. (5) they have $(l/l_d)^m$, where $l_d$ is a diffusion scale and $m > 1$ depends on the finite thickness of the last-scattering surface ($m=2$ for zero thickness ). Therefore, one can conclude that the exponential spectrum Eq. (5), described in present paper, has no relation to the photon diffusion damping. 

Let us now look to the spatial temperature maps.  Fig. 3 (adapted from the Ref. \cite{Planck}) shows a spatial CMB temperature map. Fig. 4 (adapted from the Ref. \cite{mm}) shows a spatial temperature map corresponding to the frequency spectrum shown in Fig. 1. 
It is interesting to compare (qualitatively) the fine spatial temperature chaos observed in Fig. 4 with the temperature CMB map shown in Fig. 3 (the orange region of elevated temperature near the center in the Fig. 4, corresponding to the first mode of the temperature filament, should be excluded at the comparison). 

\begin{figure}
\begin{center}
\includegraphics[width=8cm \vspace{-1cm}]{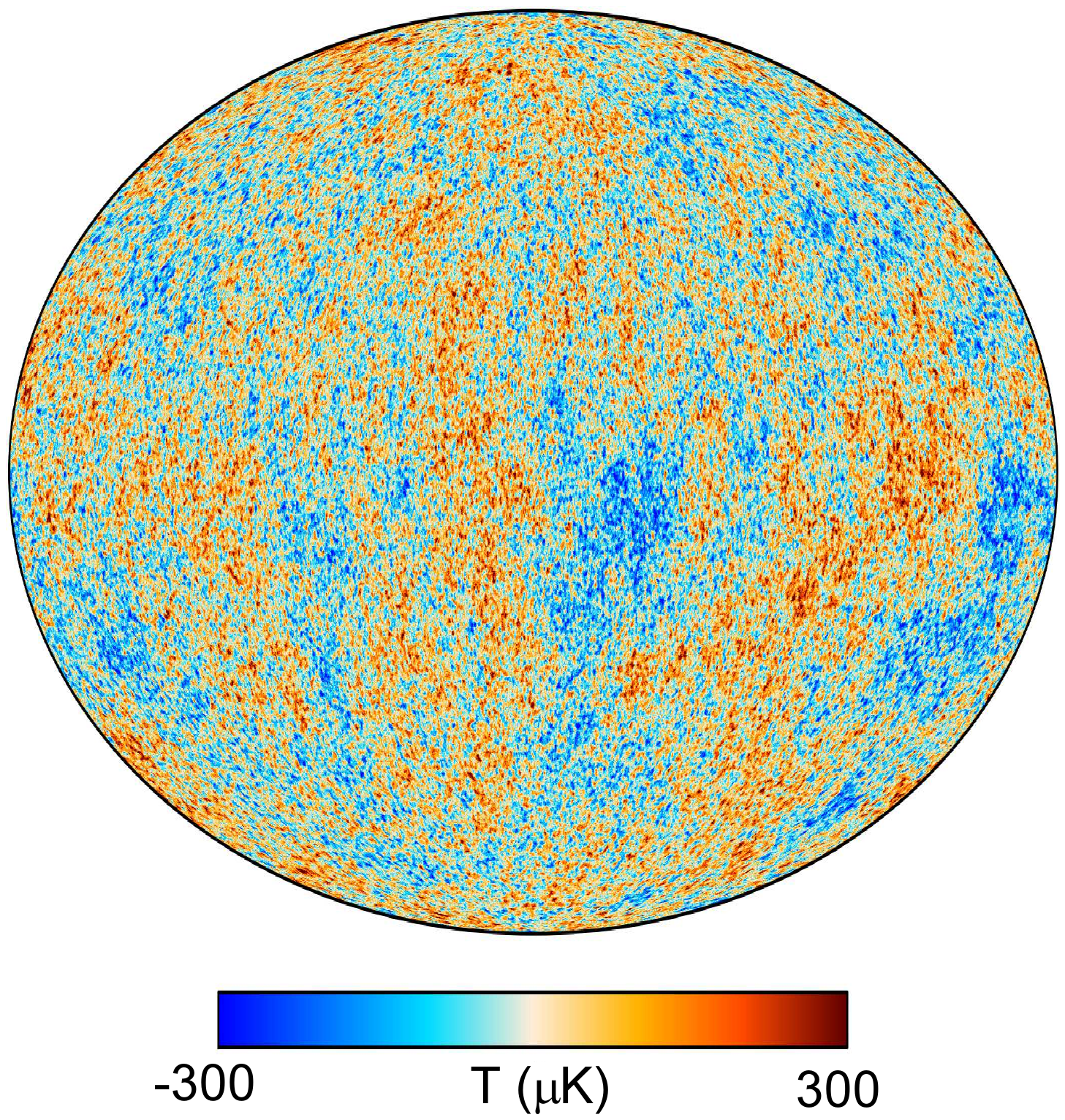}\vspace{-3.5cm}
\caption{\label{fig3}  A spatial CMB temperature map (adapted from the Ref.
\cite{Planck}). } 
\end{center}
\end{figure}

\begin{figure}
\begin{center}
\includegraphics[width=8cm \vspace{-1.3cm}]{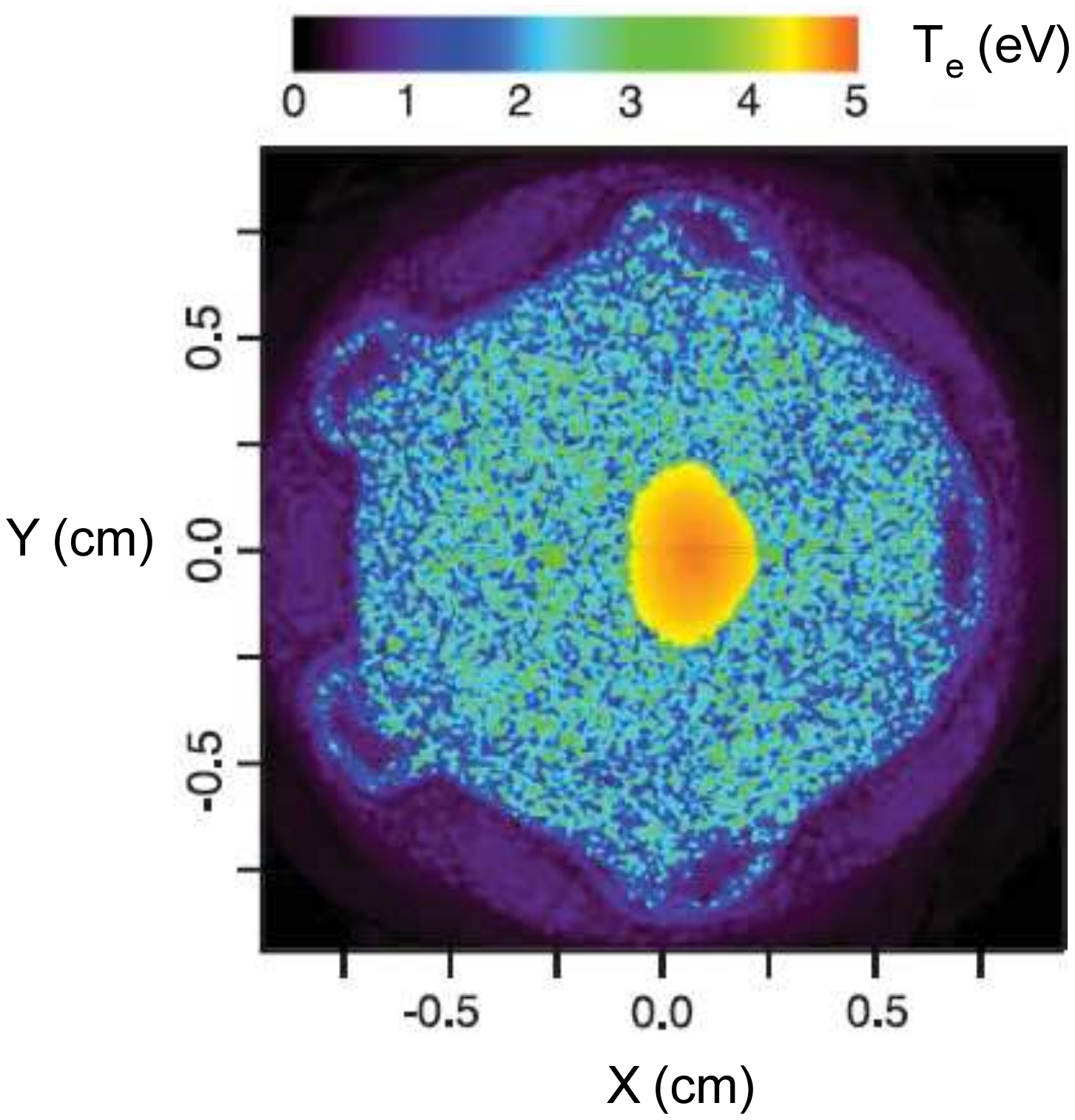}\vspace{-2.5cm}
\caption{\label{fig4} A spatial temperature map (adapted from the Ref.
\cite{mm}) corresponding to
the frequency spectrum shown in Fig. 1. } 
\end{center}
\end{figure}

\section{The 3D galaxy-galaxy power spectrum SDSS-II }
It is believed now that the large-scale inhomogeneities in the Universe were formed from initially small perturbations in the baryon-photon plasma density field at the last scattering. The first structures in the Universe were supposedly driven by the gravitational collapse of these density perturbations. The fundamental processes such as gravity, turbulence, shock waves, cooling and feedback were playing a significant role in the further structures development. At later stages, the galaxy clusters were formed. These processes are strongly nonlinear and, therefore, themselves could readily introduce chaotic dynamics. But it seems that all that were initiated by the deterministic chaos at the last scattering surface (or may be even earlier).  

\begin{figure}
\begin{center}
\includegraphics[width=12cm \vspace{-1cm}]{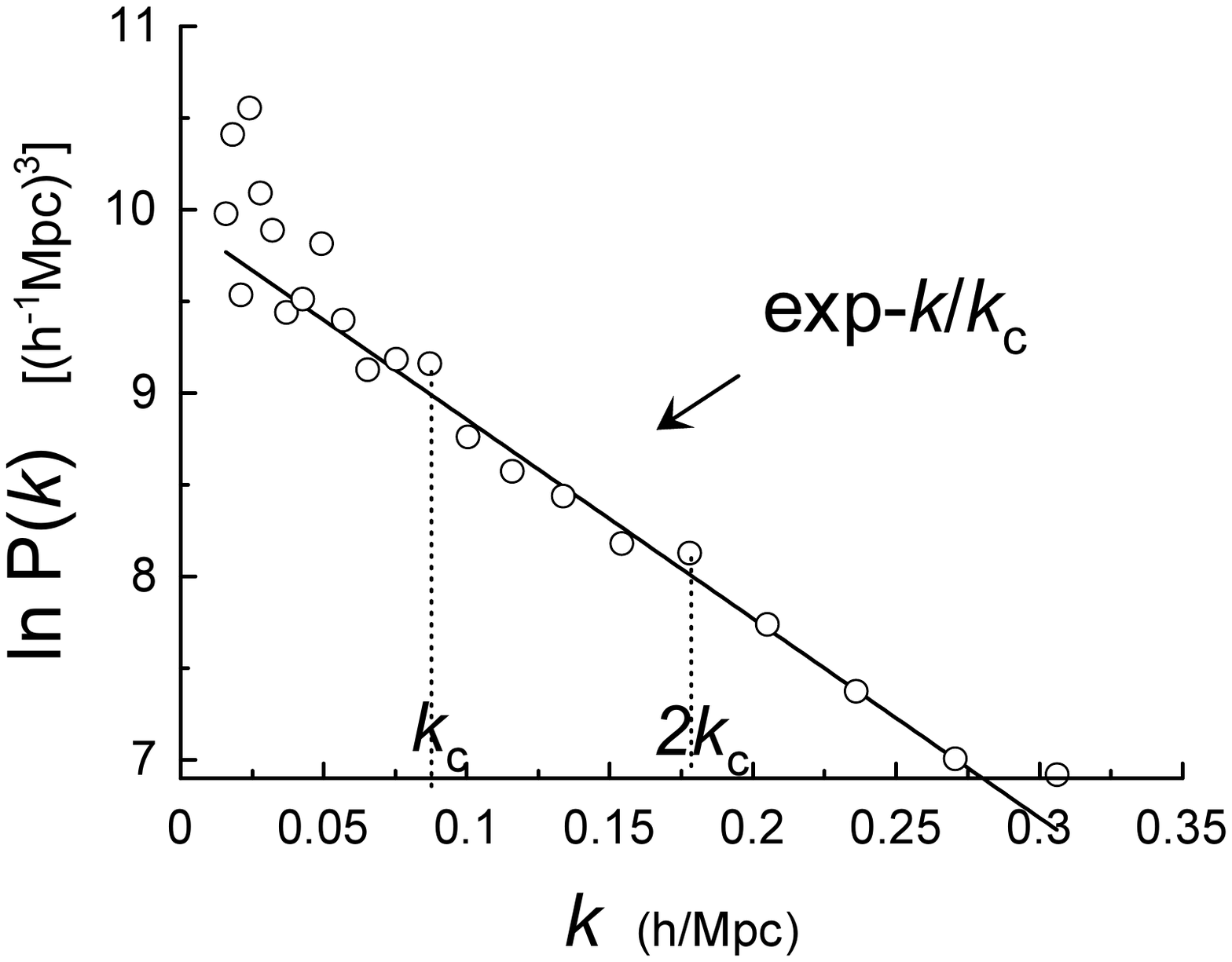}\vspace{-4cm}
\caption{\label{fig5} The power spectrum $P(k)$ (in the semi-log scales) calculated in the Ref. \cite{teg1} for the Sloan Digital Sky Survey-II. The straight line is drawn in order to indicate the exponential decay Eq. (7) with $k_c \simeq 0.09~~(h/Mpc)$.} 
\end{center}
\end{figure}

The power spectrum of a distribution of points (galaxies for a galactic sample) can be defined as the Fourier transform of the two-point correlation function:
$$
P(\bf{k})= \frac{n}{(2\pi)^{3/2}} \int \xi ({\bf r}) e^{i{\bf k}{\bf r}} d^3 {\bf r}  \eqno{(6)}
$$
where $n$ is the mean density. It seems simple but it is very difficult to calculate the real-space matter spectrum $P(k)$ from a 3D galaxy survey due to survey geometry effects, effects of light-to-mass bias, redshift-space distortions and many other practical complications. Therefore, the 3D spectrum presented in the Ref. \cite{teg1} for a sample of 205,443 galaxies from the Sloan Digital Sky Survey (SDSS-II) and covering 2417 effective square degrees (mean redshift $ z \simeq 0.1$) is rather precious one. Fig. 5 shows this (isotropic) spectrum in the semi-log scales where the exponential decay
$$
P(k) \sim \exp -(k/k_c)   \eqno{(7)}
$$
corresponds to the straight line. The waviness observed along the exponential decay has period (distance between peaks) equal to the $k_c \simeq 0.09~~(h/Mpc)$, the same $k_c$ as in the Eq. (7). It means that the waviness is generated by the same, presumably chaotic, mechanism that produced the exponential decay (cf. Introduction and Figs. 1 and 2). 

\section{Distributed chaos in the Baryon Oscillation
Spectroscopic Survey (BOSS SDSS-III) }

A more recent $P(k)$ spectrum was reported in the Ref. \cite{ander}. This spectrum was calculated using the data of the Baryon Oscillation Spectroscopic Survey (-BOSS, the largest component of the SDSS-III containing about one million galaxies). At the BOSS a technique called baryon acoustic oscillation (BAO) was used to determine the distances to galaxies (https://www.sdss3.org/surveys/boss.php.) The DR11 sample of the BOSS covers 8500 square degrees at the redshift range $0.2 < z < 0.7$. The sample discussed in the previous section was considerably smaller as well as the mean redshift $ z \simeq 0.1$. Therefore one can expect that 
the $P(k)$ spectrum calculated for the DR11 sample represents a more complex dynamics than the previous one. What can be the main difference in the dynamics? First of all it could be difference in the distributions of the characteristic wavenumbers $\kappa$ of the waves driving the chaos. In the previous case their distribution was rather narrow around $k_c$. For the more complex case the distribution can be a broad one. In this case the $P(k)$ spectrum can be represented as a weighted superposition of the exponentials:
$$
P(k) = \int_{0}^{\infty} W(\kappa ) \exp-(k/\kappa) d \kappa  \eqno{(8)} 
$$

\begin{figure}
\begin{center}
\includegraphics[width=12cm \vspace{-1cm}]{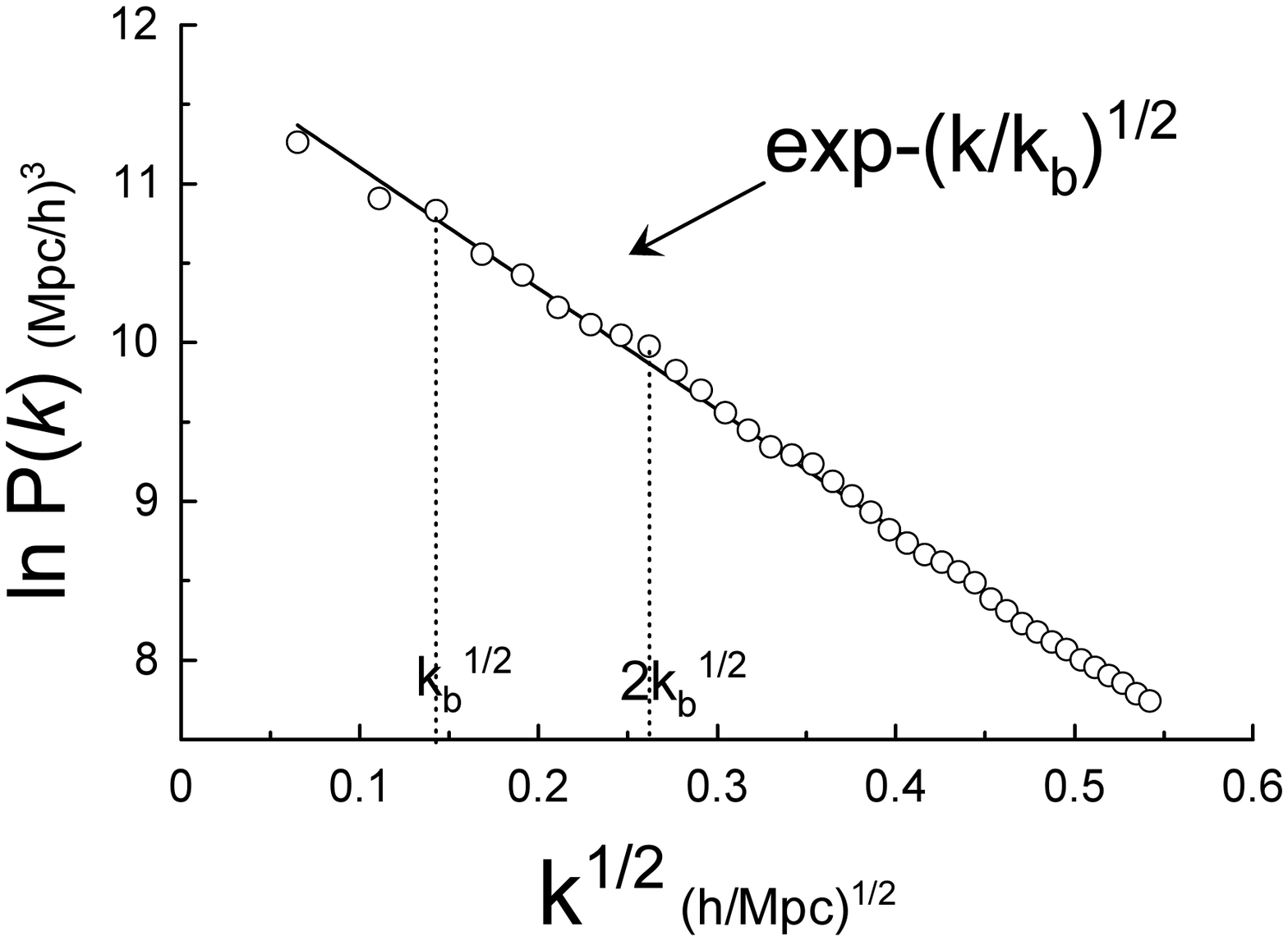}\vspace{-6cm}
\caption{\label{fig6} The power spectrum $P(k)$ calculated in the Ref. \cite{ander} for the SDSS-III BOSS DR11 (post reconstruction). The straight line is drawn in order to indicate the exponential decay Eq. (13).} 
\end{center}
\end{figure}
For the previous case the distribution $W(\kappa )$ can be approximated by delta function $W(\kappa ) \sim \delta (\kappa -k_c)$. To find a broad distribution $W(\kappa )$ for the more complex case one needs in some information about dynamics of the waves driving the chaos. Let us assume that for values of $\kappa$ for which the $W (\kappa )$ provides main contribution to the integral Eq. (8) dispersion relation of the waves exhibits scaling
$$
\omega (\kappa ) \sim   \kappa^{\alpha}  \eqno{(9)}
$$
where $\alpha$ is a scaling exponent. The concept of the surface tension ($\sigma $) is widely used in modern cosmology. At assumption that dynamics of the dispersive waves is dominated by the effects of a surface tension one can readily construct a kinematic dimension parameter controlling the kinematic relation Eq. (9). It is $\sigma/\rho $, where $\rho $ is the mass density. Then one can find the scaling exponent $\alpha$ from the simple dimension considerations:
$$  
\omega (\kappa ) \simeq a~ (\sigma/ \rho )^{1/2}~  \kappa^{3/2}  \eqno{(10)}
$$
where $a$ is a dimensionless constant. Then the group velocity
$$
v(\kappa )=\frac{d\omega}{d\kappa} \simeq \frac{3}{2}a~ (\sigma/ \rho )^{1/2}~  \kappa^{1/2}  \eqno{(11)}
$$
If the velocity has a Gaussian distribution: $\sim \exp -(v(\kappa)/v_b)^2$, then the $\kappa$ has a distribution
$$
W(\kappa ) \sim \kappa^{-1/2}~\exp -\frac{\kappa}{4k_b} \eqno{(12)}
$$
where constant $k_b = \rho v_b^2/9a^2\sigma$. Substituting the distribution Eq. (12) into integral Eq. (8) we obtain 
$$
P(k) \sim \exp-(k/k_b)^{1/2}  \eqno{(13)}
$$

Figure 6 shows the $P(k)$ reconstructed power
spectrum calculated in the Ref. \cite{ander} for the D11 sample (the data can be found at the site https://www.sdss3.org/science, look at BOSS and the file Anderson-2013-CMASSDR11-power-spectrum-post-recon-1.dat ). In the scales of the Fig. 6 
the stretched exponential dependence Eq. (13) corresponds to a straight line (with $k_b\simeq 0.017~ h/Mpc$). In this case a waviness can be  also detected along the decay with a period equal to the $k_b^{1/2}$. One can see good agreement between the data and the theoretical dependence Eq. (13). That indicates the distributed chaos and crucial role of the surface tension in the dynamics of the dispersive waves driving the chaos. \\

The author acknowledges the use of the data provided by the Planck Legacy Archive http://pla.esac.esa.int/pla and SDSS-III https://www.sdss3.org/science. I also grateful to A. Shelest for attracting my attention to the data.

\end{document}